\documentclass[
reprint,
aps,
 amsmath,amssymb,
 prf,twocolumn,
]{revtex4-1}

\usepackage[utf8]{inputenc}
\usepackage{physics}
\usepackage{graphicx}% Include figure files
\usepackage{bm}% bold math
\usepackage[colorlinks=true, allcolors=blue]{hyperref}

\newcommand\Nd{\mathcal{N}(d)}
\newcommand\Rey{\mathrm{Re}}
\newcommand\We{\mathrm{We}}
\newcommand\lic{\ell_{ic}}

\begin{document}

\title{Breakup cascade in gas filament}

\author{Ali\'enor Rivi\`ere$^{1,2}$}
\email{alienor.riviere@epfl.ch}
\author{Zehua Liu$^3$}
\author{Jishen Zhang$^2$}
\author{Laurent Duchemin$^2$}
\author{Luc Deike$^{3,4}$}
\author{St\'ephane Perrard$^2$}

\affiliation{$^1$LFMI, \'Ecole Polytechnique F\'ed\'erale de Lausanne, CH-1015 Lausanne, Switzerland}
\affiliation{$^2$PMMH, CNRS, ESPCI Paris, Universit\'e PSL, Sorbonne Universit\'e, Universit\'e de Paris, F-75005, Paris,
France}
\affiliation{$^3$Department of Mechanical and Aerospace Engineering, Princeton University, Princeton, NJ 08544, USA}
\affiliation{$^4$High Meadows Environmental Institute, Princeton University, Princeton, NJ 08544, USA}


\begin{abstract}
Despite its importance in both geophysical and industrial contexts, the inertial fragmentation of gas filaments has received much less attention than their liquid counterparts. 
Yet, gas filaments produce the smallest bubble sizes, which drive gas dissolution, critical to ocean-atmosphere exchange such as carbon dioxide and oxygen, as well as marine aerosols emission, serving as nuclei for cloud condensation and ice particle production.
Here, we unravel the fundamental physics governing the splitting of a single filament in a model geometry by combining numerical simulations, laboratory experiments and theory.
We show that the splitting of a single filament generates a power-law bubble size distribution following $d^{-3/2}$ with $d$ the volume equivalent bubble diameter, suggesting the existence of a self-similar breakup mechanism, absent in liquid ligament fragmentation.
We propose a deterministic model, based on the capillary fragmentation of a filament with power-law shape, which quantitatively captures the bubble size distribution.
We demonstrate that the filament shape at breakup sets the size distribution of a first generation of bubbles. This distribution is then reproduced at smaller and smaller scales by latter breakups in a self-similar manner.
The $d^{-3/2}$-distribution coincides with the size distribution of small bubbles observed in dilute turbulent flow, such as below breaking waves. We argue that the turbulent bubble size distribution observed in nature arises as the superposition of many individual filament splittings. The turbulence nature of the flow only sets the initial conditions of each splitting dynamics, and play no role in the bubble size selection.
\end{abstract}

\maketitle
\section{Introduction}
Bubbles significantly enhance momentum, mass, and chemical transfers in a wide range of industrial and environmental contexts, including bubble column reactors \cite{han2007,kantarci2005,risso2018}, emulsifiers, ocean surfaces \cite{liang2011,deike2022,jiang2022}, rivers \cite{beaulieu2012}, and waterfalls \cite{demars2013}.
All these transfers depend critically on the bubble size, hence playing a crucial role in industrial processes optimization or in the modeling of geophysical mechanisms. The quantity of interest is the bubble size distribution, $\Nd$, with $d$ the volume-equivalent bubble diameter.

In dilute turbulent environments, such as below breaking waves, the bubble size distribution exhibits two power-law scalings \cite{deane2002,riviere2022}, separated by the size at which inertial forces balance capillarity, called the Kolmogorov-Hinze scale $d_h$ \cite{kolmogorov1949,hinze1955}.
Power-law distributions are generally associated to scale-invariant physical processes.
The length slide of avalanches \cite{faillettaz2004,brunetti2009}, the intensity of earthquakes \cite{meng2019} or the energy distribution among scales in fully developed turbulence \cite{K41} are ones over many examples of scale invariant processes.
%This size separates statistically stable ($d<d_h$) from unstable ($d>d_h$) bubbles.
The super-Hinze ($d>d_h$) bubble size distribution was shown to emerge from a scale invariant process, namely a self-similar breakup process occurring on a time controlled by the turbulent flow surrounding each bubble \cite{garrett2000}.
%In contrast, the self-similar process leading to the sub-Hinze bubble size distribution remains to be identified.
In previous works \cite{riviere2022,ruth2022}, we identified that sub-Hinze bubbles ($d<d_h$) are produced via the capillary splitting of air filaments produced by large super-Hinze bubble deformations in turbulence.
Yet, the physical process driving filament fragmentation, which leads to a power-law size distribution was not identified.

Noteworthy, the symmetric scenario of liquid ligament fragmentation in air \cite{villermaux2004,villermaux2007,eggers2008,villermaux2012} does not exhibit power-law drop size distribution.
Instead, the size distribution arising from the breakup of smooth liquid ligament is peaked around specific length scales selected by instabilities~\cite{eggers2008, villermaux2020}.
Small amplitude noise on the initial conditions widens the distribution around these selected sizes but the overall shape remains the same \cite{pal2024}.
In liquid jets and sprays, ligaments are highly corrugated and several modes sum up to produce the final drop size distribution \cite{villermaux2020}.
The discrepancy between liquid and gas breakup is not limited to inertial fragmentation.
When approaching breakup, the neck dynamics of drops follows universal exponents and a self-similar behavior which contrasts with the faster non universal bubble pinch-off dynamics \cite{suryo2004,keim2006,bergmann2006,eggers2007,thoroddsen2007,thoroddsen2008,pahlavan2019}.
Similarly, in micro-fluidic devices, long liquid ligaments are widely used to produce monodisperse drops when special devices must be used to produce monodisperse bubbles owing to the absolute instability of gas filaments \cite{gordillo2001,ganan2006,castro2012,fu2015}.
Hence, to our knowledge, only one study, by Thoroddsen and coauthors, describes gas filament fragmentation following bubble breakup \cite{thoroddsen2007}.
They observed that a bubble in a viscous quiescent fluid detaching from a needle tip
can create a long gas thread, which then destabilizes forming dozens of micron-size child bubbles. The bubble size distribution produced was however not quantified.

In this work, we seek to identify the physical process controlling the fragmentation of stretched filaments.
We perform simulations of a single filament in an extensional flow, combined with breakup experiments in a similar model flow geometry.

\section*{Fragmentation of a stretched filament}

\begin{figure*}[!ht]%[!tbhp]
\centering
\includegraphics{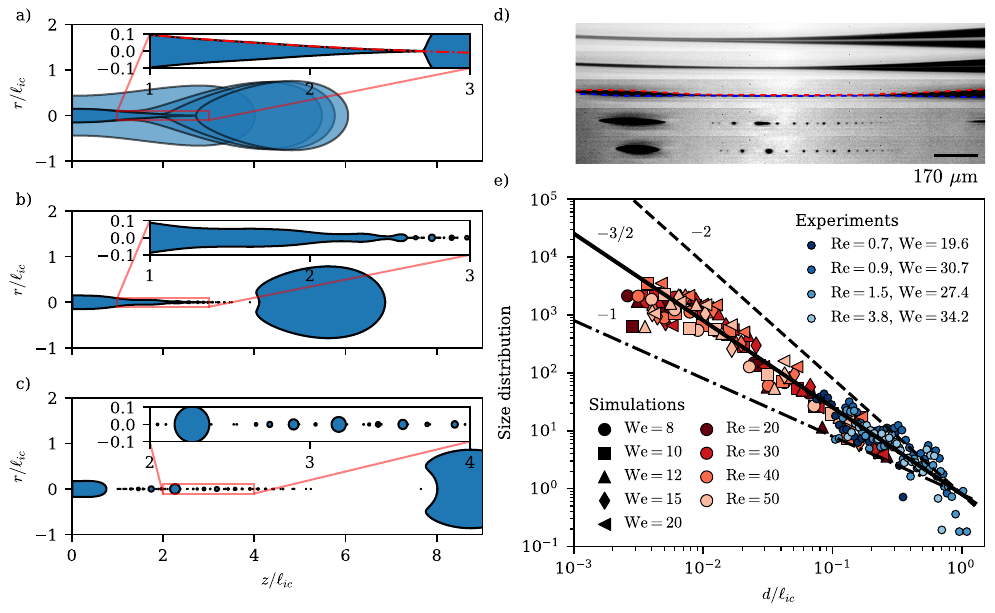}
\caption{Filament generation and fragmentation. a-c) Typical deformation and fragmentation sequence obtained in the numerical simulation, here at $\Rey=40$, $\We=15$. Shapes are plotted at $S(t - t_b) = -0.3, -0.08, -0.04, 0, 0.1$ and $ 0.4$, respectively where $t_b$ is the breakup time, and $1/S$ the large-scale flow timescale. a)~The initial bubble deforms under the action of the flow creating an elongated filament of size a few $\lic$. The inset shows an enlargement at pinch-off around the pinch-off point. The red dashed line is a polynomial fit of degree 2. b)~The filament shape destabilizes under capillary forces, visualized in the enlargement, and breaks into dozens of bubbles. c)~Small bubbles are produced in between larger ones, leading to a non monotonic sequence of bubbles sizes in space, and a very broad range of bubble sizes, small compared to $\lic$.
d)~Experimental time sequence showing gas filament rupture and subsequent bubble formation, at $\mathrm{Re}=3$, $\mathrm{We}=23$ with a time interval of $161~\mathrm{\mu s}$ between frames, corresponding to $0.08 d_\text{needle}/U_\text{needle}$, the large scale time scale. Red and blue dashed lines are polynomial fits of degree two on the upper (resp. lower) filament shape.
e)~Experimental and numerical bubble size distributions, normalized such that their integrals equals the number of bubbles produced. Experimental curves are vertically shifted by a factor 5 to match the numerics. While experimental curves are averaged over hundreds of experiments, numerical distributions results from the fragmentation of individual filaments at various Reynolds and Weber numbers, encoded by the marker color and shape, respectively. Sizes are in units of the inertio-capillary length scale $\lic$. All data sets show a clear $d^{-3/2}$-scaling over several orders of magnitude in bubble sizes.}
\label{fig:intro}
\end{figure*}

We perform numerical simulations and laboratory experiments mimicking the deformation and fragmentation of a bubble of negligible viscosity and density compared to the surrounding fluid in a straining flow (see methods for details of the experimental and numerical setups). The liquid density and kinematic viscosity are $\rho$ and $\nu$, respectively, the surface tension between the liquid and the gas is $\gamma$. The dynamics depends on two dimensionless numbers, the Reynolds number $\Rey = Ud/\nu$, which compares inertial and viscous forces, and the Weber number $\We = \rho U^2 d/\gamma$, which compares inertial and capillary forces, where $U$ and $d$ are a characteristic velocity and a length scale, respectively. In the numerical simulation, $d=d_0$, the initial bubble diameter and $U=Sd_0$, with $S$ the imposed strain of the flow. In the experiment, $d=d_{\text{needle}}$ the needle diameter, and $U=U_{\text{needle}}$ the initial velocity of the needle.
Deformation and fragmentation being primarily controlled by the competition between inertia and capillarity, length are discussed in units of the inertio-capillary length scale $\lic$, defined by $\mathrm{We}(\lic) =1$ (see methods) in both the numerical simulation and the experiment.
For scales large compared to the inertio-capillary length, the surrounding flow influences the deformation dynamics, while it becomes negligible for scales smaller than $\lic$.

Fig.~\ref{fig:intro}a-c show a typical sequence of deformation and fragmentation observed in the simulation, here at $\Rey = 40$ and $\We =  15$.
The deformations prior to the first breakup are illustrated on fig.~\ref{fig:intro}a, for four dimensionless times $S(t - t_b)$, where $t_b$ is the breakup time.
The initially spherical bubble first deforms into a cigare shape pulled by the flow along the axial direction and compressed along the radial direction ($S(t-t_b) = -0.3$). The bubble radius shrinks until the formation of an elongated central region, which we call a filament, and a head (visible at $S(t-t_b) = -0.08$).
Once a local minimum appears on the filament radius, most deformations are concentrated in a small region surrounding the minimum radius, whose location drifts in time along the axis of symmetry. At pinch-off, the minimum radius reaches zero, creating a thin filament and two child bubbles (by symmetry) of size comparable to the initial bubble size ($S(t-t_b) = 0$).
The filament shape at breakup is polynomial of degree two (red dashed line in the inset) and extends over several inertio-capillary lengths.
The filament then destabilizes under capillary effects (fig.~\ref{fig:intro}b) creating a large number of child bubbles smaller than the inertio-capillary lengths (inset plot and fig.~\ref{fig:intro}c).
The breakup of the filament forms a perl chain of bubbles of heterogeneous sizes. The bubble sizes are not ordered in space: tiny bubbles compared to $\lic$ are often interspersed between larger ones.
Consequently, child bubble sizes span over several orders of magnitude, similar to the observation from Thoroddsen et al. (2007) \cite{thoroddsen2007}.
A similar splitting process is observed experimentally (fig~\ref{fig:intro}d), leading to the production of dozens of micron-size bubbles (here at $\We = 23$, $\Rey = 3$).

Next, fig.~\ref{fig:intro}e shows the bubble size distribution generated by the breakup of gas filaments in both numerical simulations and experiments. Experimental distributions are averaged over hundreds of repeated runs and show a power-law distribution compatible with $d^{-3/2}$ over one order of magnitude. In simulations, each curve corresponds to the splitting of a single filament at a given Reynolds and Weber number. Each distribution shows a clear $d^{-3/2}$-power law, spanning over two orders of magnitude in bubble sizes, the largest bubble produced being of size $0.5 \lic$.
Hence, a power law distribution for the bubble size emerges from the fragmentation of a single filament, in a uniaxial strain flow.
The Weber and the Reynolds numbers affect the total filament volume (See Fig. S4, S5, S6 in SI Appendix), but not the fragmentation process itself.

Notably, the $d^{-3/2}$-power-law bubble size distribution coincides with the one observed in a turbulent flow for sub-Hinze bubbles.
This fundamental result suggests that the turbulent bubble size distribution arises as the sum of individual filament splitting, all generating a $d^{-3/2}$-distribution.
We note that in turbulence, bubbles tend to align with the principal straining direction \cite{masuk2021}. When they break, as noticed in previous works~\cite{rodriguez2006,revuelta2006,gordillo2005}, the surrounding flow geometry is typically an axi-symmetric uniaxial straining flow which inspired the geometry used here.

\section*{A capillary fragmentation cascade}
\begin{figure}
\centering
\includegraphics{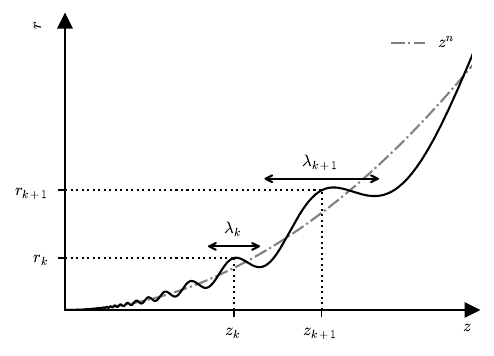}
\caption{Scheme of the filament shape at the time of the initial breakup envisioned in our model. The overall filament shape follows a power-law scaling $z^n$, on which capillary waves, of wavelength $\lambda(r) \sim r$, grow. The local wavelength, $\lambda_k$, will give birth to the $k^{th}$-child bubble located at position $z_k$.}
\label{fig:scheme}
\end{figure}

In this section, all lengths are in units of the inertio-capillary length.
To rationalize the origin of the bubble size distribution, we envision the following scenario, schematized on figure~\ref{fig:scheme}: (i)~the initial stretching prepares a filament shape $r\sim z^n$, (ii)~unstable capillary waves, of local wavelength $\lambda(r)\sim r$, grow on the surface of the filament, (iii) the filament splits into numerous child bubbles, the volume of the $k^{th}$ child bubble being $d_k^3 \sim \lambda_k r_k^2$.
Next, the distance between consecutive bubbles relates to the local wavelengths through $z_{k+1} - z_{k} = (\lambda_{k+1} + \lambda_k)/2 $. This last relation can be rewritten using (i) and (ii) in terms of filament radii only $r_{k+1}^{1/n} - r_{k}^{1/n} \sim r_{k+1} + r_k$. Then, using (ii) and (iii) we get a scaling for the $k^{th}$ bubble size $d_k \sim r_k$ which leads to the following sequence of bubble sizes
\begin{equation}
    \frac{d_{k+1}^{1/n} - d_k^{1/n}}{d_{k+1} + d_k} = C_n
    \label{eq:Bubsequence}
\end{equation}
with $C_n$ a constant of order 1 which might depend on the surrounding flow field and therefore on both $\Rey$ and $\We$.
Notably, under continuous assumptions, the previous set of hypothesis implies a power-law scaling for the bubble size distribution $\Nd = \dd k / \dd d$ where $\dd k$ is an elementary number of bubbles.
Indeed, in a continuous setting, the geometrical constraint rewrites $\dd z/\dd k \propto \lambda $. Taking the inverse, and substituting $\lambda$ by $z^{n}$, we obtain the number density of bubbles per unit length  $\dd k/\dd z \sim z^{-n}$. Then, using $d \sim r$ and (i) we get $\dd d/\dd z\sim z^{n-1} $
By chain rule, $\Nd = \dd k/\dd d = \dd k/\dd z\times \dd z/\dd d \propto z^{-2n +1}$, which gives, after substitution of $z$ by $d^{1/n}$,
\begin{equation}
    \Nd \propto d^{[-2n+1]/n}.
\end{equation}
Hence, the filament shape controls the power-law of the bubble size distribution. In particular, $n=2$, corresponding to a quadratic filament shape, gives $\Nd \propto d^{-3/2}$, the sub-Hinze bubble size distribution.

\section*{A size distribution controlled by the initial filament shape}
\subsection*{Child bubbles identification}
\begin{figure*}[!t]
\centering
\includegraphics{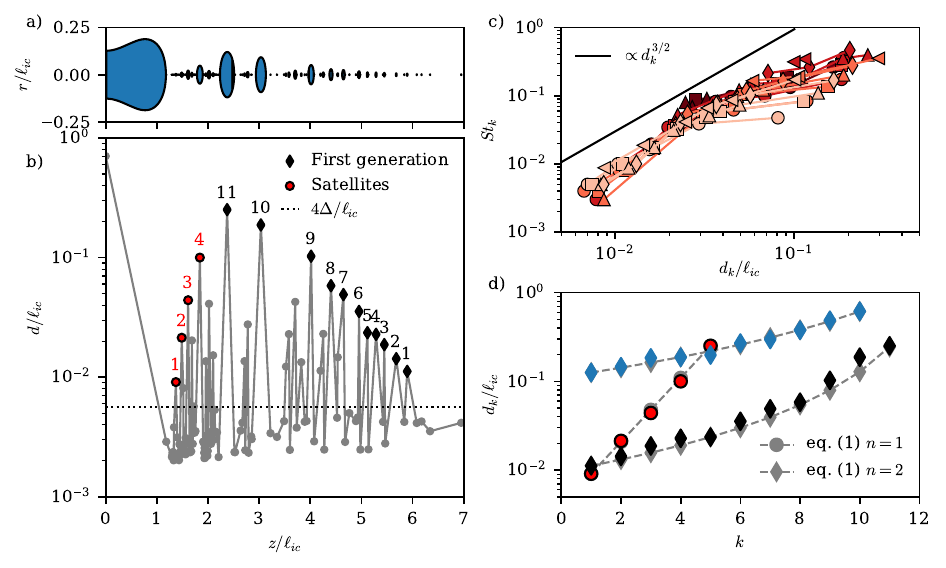}
\caption{Size sequence of bubble sizes within the filament. a)~Example of child bubbles produced at the end of the fragmentation process of a filament in a simulation at $\Rey=30$, $\We=12$ and their corresponding sizes (b). b)~We identify two types of child bubbles: 1. \textit{first generation} bubbles resulting from the splitting of the overall filament shape, denoted by black diamonds. 2. \textit{Satellites} produced by the subsequent splitting of first generation bubbles. Satellites produced by the splitting of the largest child bubble (11) are denoted by red circles. Bubbles of size below 4 grid points, $d<4 \Delta$, are considered unresolved. 
c)~Breakup time of first generation bubbles, $t_k$, measured in the simulation in units of the flow timescale $1/S$, as a function of their size $d_k$. See figure~\ref{fig:intro} for color and symbols' meaning. All curves collapse on the same master curve, compatible with a $d_k^{3/2}$-scaling corresponding to the capillary timescale on a filament of diameter $d_k$.
d)~Examples of bubble size sequence in an experiment at $\mathrm{Re}=3.8$, $\mathrm{We}=35.1$ (blue symbols), and a simulation at $\Rey=30$, $\We=12$.
First generation bubbles are denoted by diamonds and satellites by circles.
Grey lines are the prediction from model~\eqref{eq:Bubsequence}, using $n=1$ for satellite bubbles and $n=2$ for first generation bubbles and built sequentially starting from the smallest bubble size within each filament.
The coefficient $C_n$ is evaluated for each filament by averaging \eqref{eq:Bubsequence} over all pairs ($d_k$, $d_{k+1}$) within the filament.
The agreement between the prediction and the size sequence is excellent.}
\label{fig:proof}    
\end{figure*}

To test the validity of the model, we now look at the spatial organization of bubbles generated by the sequence of fragmentation.
Fig.~\ref{fig:proof}a shows a typical snapshot of the numerous child bubbles produced by the filament splitting, for $\Rey = 30$ and $\We = 12$.
Fig.~\ref{fig:proof}b shows the corresponding bubble sizes $d$ as a function of their position $z$, evidencing a complex spatial pattern. In this sequence, we identify the local largest bubbles produced (black diamonds), forming an envelop reminiscent of the initial filament shape. These \textit{first generation} bubbles are produced sequentially, and are then numbered by their time of production.
Multiple cascades of fragmentation then occur concomitantly, in between each adjacent pair of first generation bubble, generating smaller and smaller \textit{satellites} bubbles. The largest local satellites produced by the splitting of the largest child bubble of first generation (child 11) are denoted by red circles, and are also numbered by their time of production. When satellites bubbles are produced, the filament is not stretched anymore and we expect the initial flow to have only little influence on their production.
We run complementary simulations, starting from the filament shape at breakup with zero velocity everywhere. We find that the same bubble size distribution emerges (see SI Appendic fig. S3): the initial flow prepares a filament shape, the following cascade of events is then fully independent of the surrounding flow.

\subsection*{Capillary breakup time and size sequence}

We now test the model validity on the first generation bubble and a subset of the satellite bubbles (red circles), for all Weber and Reynolds number. Figure~\ref{fig:proof}c shows the breakup time $t_k$ associated to the production of each \textit{first generation} bubble, measured in the simulation, as a function of their size in units of $\lic$. Time is in units of $1/S$. Notably, first generation bubbles appear sequentially, the smallest bubble being produced first, in a mechanism resembling the successive bubble production in a receding gas bridge
\cite{duchemin2003}.
We find that all data collapse on a master curve, following a $d_k^{3/2}$-scaling (solid black line).
Deviations are observed for the largest bubble in each dataset. We attribute their faster breakup to the presence of the surrounding flow which might accelerate breakup at these scales.
The scaling for the breakup time corresponds to the growth rate of capillary modes onto a filament of local radius $\sim d_k$, and has also been observed for bubble breakup in turbulence~\cite{riviere2022}. This result confirms that splittings are capillary-driven.

For each simulation ($\Rey, \We$), we then test the size sequence by measuring the ratio~\eqref{eq:Bubsequence} for all consecutive bubbles, considering separately first generation and satellite bubbles. 
The model is considered valid if, for a given $n$, the ratio $C_n$ is approximately constant within the size sequence.
We tested different integers values for $n$. We denote by $C^{\text{fg}}_n$ and $C^{\text{s}}_n$, the ratios evaluated for first generation and satellite bubbles, respectively.
Our analysis reveals that, for bubbles of \textit{first generation}, the ratio $C^{\text{fg}}_n$, is constant along each filament when $n=2$, while it shows a decreasing trend for $n=1$ and an increasing trend for $n=3$ (shown in SI fig. S7).
Fig.~\ref{fig:proof}d compares first generation bubbles sizes as a function of $k$, in both a simulation at $\Rey = 30$, $\We = 12$ (black diamonds), and an experiment at $\Rey = 3.8$, $\We = 35.1$ (blue diamonds), with the prediction of our model~\eqref{eq:Bubsequence} with $n=2$ (whose solution in given in Eq. 4 in SI).
The model prediction is built sequentially starting from the smallest bubble size, using the averaged value of $C^{\text{fg}}_2$, evaluated by averaging over all the pairs ($d_k^{\text{fg}}, d_{k+1}^{\text{fg}}$) within the filament, with no adjustable parameter.
In both cases, the agreement is excellent.
Other examples can be found in SI Appendix (fig. S10), all showing an excellent agreement.
The average value of $C^{\text{fg}}_2$ is evaluated in each simulation, and is of order 1 regardless of the values of $\Rey$ and $\We$ (see SI fig. S9).
The same analysis performed for \textit{satellite} bubbles reveals that a linear profile, $n=1$, must be chosen instead, as a clear decreasing trend is found for $n\geq 2$ (see SI fig. S8).
The linear filament profile implies a proportional relationship between consecutive bubble sizes $d_{k+1}^\text{s} = (C_1^\text{s} +1 )/(C_1^\text{s}-1)d_k^\text{s}$.
Figure~\ref{fig:proof}d compares the prediction from~\eqref{eq:Bubsequence}, with $n=1$ and $C_1^\text{s}$ evaluated as the average ratio, with satellites' sizes in a simulation at $\Rey = 40$, $\We = 15$.
Again, the agreement is excellent (see SI fig. S11 for other examples).
The constant $C_1^\text{s}$ is again of order one, showing no dependence on neither $\Rey$ nor $\We$ (see SI fig. S9) suggesting that the flow does not influence the subsequent filament splitting that produces satellite bubbles.
A similar linear relationship between satellite drop diameters was previously found in
Tjahjadi et al. (1992)~\cite{tjahjadi1992} for drop fragmentation cascade in a viscous quiescent fluid.

The two different values of $n$ can be understood by comparing the filament length to the inertio-capillary length $\lic$. If the filament length is small compared to $\lic$, then the outer fluid inertia is negligible and the linear profile close to pinch-off controls the satellite size selection \cite{thoroddsen2007,eggers2007}, coherent with the $n=1$ regime.
Conversely, when the filament length is large compared to $\lic$, as it is the case for the initial filament shape, then the outer fluid inertia dominates the dynamics and the overall neck shape prior to breakup is quadratic \cite{burton2005,thoroddsen2007,burton2008}.

To summarize, the $d^{-3/2}$-distribution arises as the sum of two physical processes: first, stretching creates a filament shape $r\sim z^2$ which fragments producing bubble sizes following a $d^{-3/2}$-distribution. Second, these bubbles break, generating satellites bubbles of size proportional to the first generation. This linear relation between first generation and satellite bubble sizes implies that the $d^{-3/2}$-distribution is reproduced at smaller and smaller scales.
This two-stages cascade scenario is very different from breakups observed in liquid ligaments. When the surrounding fluid has negligible inertia and viscosity, the ligament breakup is controlled by hydrodynamics instabilities. These instabilities select a wavelength and therefore a discrete set of sizes corresponding to the most unstable wavelength in the system.
Initial roughness on the ligament shape \cite{eggers2008} widen the distribution around this most unstable wavelength through coalescence of corrugations.
Hence, the overall drop size distribution appears as the convolution between the ligament radius distribution and corrugation of individual ligaments.

By investigating the successive fragmentation stages of a filament, we connect the classic fragmentation picture of a bubble breaking into $N$ child bubbles, with the statistical analysis of elementary binary breakup events of a bubble immersed in a turbulent flow proposed by Rivi\`ere et al. \cite{riviere2022}. They showed that the production time of small bubbles corresponds to the capillary timescale defined at the filament scale, leading to a $d^{-3/2}$-size distribution. Here we show that capillarity drives both the production time and the size selection through the filament shape, and we identified \textit{how} these capillary controlled child bubbles were individually produced.

\section*{Conclusion}
By performing numerical simulations and experiments of a single filament splitting, we demonstrate that the splitting of a single filament generates a power-law bubble size distribution.
We present a deterministic model, based on the capillary fragmentation of an axi-symmetric filament of power-law shape which quantitatively reproduces bubbles sizes and explains the power-law scaling of the size distribution.
In the presence of a surrounding strain, the flow stretches the filament into a parabolic shape, which then controls the power-law of the bubble size distribution.
This shape sets the size distribution of first generation bubbles following $d^{-3/2}$.
Latter breakups occur at scales much smaller than the inertio-capillary length and are therefore unsensitive to the presence of the initial strain flow.
The later child bubble generation size being proportional to the first generation bubble sizes, the $d^{-3/2}$-distribution is copied at smaller and smaller scales through a self-similar process. 

The self-similar breakup mechanism should produce bubbles of arbitrary small size, which might explain the challenges in microfluidics devices aiming at producing single sized small bubbles. Separately, our theoretical model suggests that the ability to produce filaments of any power-law shape would lead to the control of the subsequent bubble size distribution, with could be leveraged in micro-fluidic devices.

The emerging $d^{-3/2}$-distribution of bubble sizes coincides with that observed for sub-Hinze bubbles in a turbulent flow, suggesting that single filament splitting controls the final size distribution.
The surrounding turbulent flow prepares the initial filament shape, but plays no role in the fragmentation process itself, similarly to bubble pinch-off in turbulence \cite{ruth2019}.
In a turbulent flow, the sub-Hinze distribution arises as the sum of the size distributions generated by the individual deterministic filament splitting.

\section{Material and Methods}
\subsection*{Numerical simulations}
We simulate the deformation and breakup dynamic of a bubble fixed at the center of a uniaxial straining flow of velocity field written in cylindrical coordinates (r, z): $u_r = -1/2Sr$, $u_z = Sz$, with $S$ the characteristic strain.
The numerical simulations are performed using the open-source software Basilisk \cite{popinet2003,popinet2009}, which uses finite-volume schemes and a Volume-Of-Fluid method, combined with a sharp interface reconstruction, to solve the two-phase incompressible Navier-Stokes equations.
The solver has been extensively validated in the past to tackle complex capillary driven processes, such as splashing \cite{marcotte2019,thale2024}, bubble bursting \cite{berny2021} or liquid rim fragmentation \cite{tang2024}.
We solve the two-phase axi-symmetric Navier-Stokes equations on a square domain of size $L=10R_0$, with $R_0$ the initial bubble radius, with imposed symmetry with respect to the plane $z=0$.
Density and dynamic viscosity ratios are chosen close to the air-water ratios, $\rho_R = 850$ and $\mu_R = 25$, respectively. We define the inertio-capillary length in the simulation by $\lic = [\gamma/\rho S^2]^{1/3}$.
The simulation goes in two steps. First a uniaxial straining flow at a given $\Rey$ is created by imposing the adequate boundary conditions: Dirichlet boundary conditions at inflow, and Neumann boundary conditions at the outflow for the velocity and the contrary for pressure.
The flow converges to the theoretical flow on a few characteristic timescales $1/S$.
Then the spherical bubble is injected at the center of the stagnation point flow by changing locally the density and viscosity.
The Weber number is changed by adapting the surface tension coefficient.
We extensively make use of the adaptive meshgrid refinement to capture the thin scales of the filament and the very small bubbles, while saving computational time. The minimum grid size $\Delta$ is such that $R_0/\Delta \approx 820$. This numerical set-up has been previously validated to describe bubble deformations in a uniaxial straining flow~\cite{riviere2023}.
Numerical convergence of the bubble size distribution is verified by changing the smallest grid size (see SI fig S2).

\subsection*{Experimental setup}
The filament pinch-off is investigated experimentally (sketch of the experimental set-up in SI fig. S1). A movable needle with a diameter of $d_\text{needle}=1.2$~mm, closed at one end, is initially submerged horizontally in a liquid bath filled with silicone oil. The open end of the needle is connected to an integrated air reservoir, enabling the attachment of an air bubble to the needle tip. To initiate the experiment, a mass is released from a given height, striking the tip of an "L-shaped" bellcrank arm. This impact induces the rapid rotation of the arm, causing the needle to be withdrawn horizontally from the liquid bath. As a result, a gas filament forms, stretches and subsequently ruptures. The breakup events are recorded at 6200 Hz using a Phantom v1840 high-speed camera equipped with a Navitar optical lens leading to a pixel size of 0.9 $\mu$m at best. The horizontal velocity of the needle at the rupture instant is subsequently determined from the image sequences.
The experimental Reynolds number ($\Rey$) and Weber number ($\We$) are defined using the needle diameter $D$, withdrawal speed $U_\text{needle}$, liquid viscosity $\nu_s$, and surface tension $\gamma_s$. Silicone oils with viscosities of 250, 500, and 1000 cSt are employed to independently vary $\Rey$ and $\We$. We define the inertio-capillary length by $\lic = \gamma /(\rho U_\text{needle}^2)$ in the experimental set-up.

\begin{acknowledgments}
The authors thank François P\'etr\'elis for insightful scientific discussions on the theoretical model and Laurent Quartier for help in the experimental setup. This work was supported by NSF grant 2242512 to L.Deike. This work was funded by a PSL Starting Grant 2022, as well as the ANR Lascaturb (reference ANR-23-CE30-0043-03) to S. Perrard.
\end{acknowledgments}

% Bibliography
\bibliography{biblio}

\end{document}